\documentclass[%
 reprint,
 amsmath,amssymb,
 aps,
]{revtex4-2}

\usepackage{graphicx}% Include figure files
\usepackage{dcolumn}% Align table columns on decimal point
\usepackage{bm}% bold math
\usepackage{hyperref}
\usepackage[dvipsnames]{xcolor}

\begin{document}

\preprint{APS/123-QED}

%\title{Computer-assisted pattern recognition of quasicrystalline surfaces}
\title{Automatic determination of quasicrystalline patterns from microscopy images}

\author{Tano Kim Kender}
    \affiliation{Faculty of Physics and Center for Computational Materials Science, University of Vienna}
    
\author{Marco Corrias}
\email{marco.corrias@univie.ac.at}
    \affiliation{Faculty of Physics and Center for Computational Materials Science, University of Vienna}
    
\author{Cesare Franchini}
\email{cesare.franchini@univie.ac.at}
    \affiliation{Department of Physics and Astronomy, University of Bologna}
    \affiliation{Faculty of Physics and Center for Computational Materials Science, University of Vienna}

\date{\today}

\begin{abstract}
%Since their discovery, quasicrystalline systems have been extensively studied by the scientific community.
Quasicrystals are aperiodically ordered solids that exhibit long-range order without translational periodicity, bridging the gap between crystalline and amorphous materials.
Due to their lack of translational periodicity, information on atomic arrangements in quasicrystals cannot be extracted by current crystalline lattice recognition softwares. This work introduces a method to automatically detect quasicrystalline atomic arrangements and tiling using image feature recognition coupled with machine learning, tailored towards quasiperiodic tilings with 8-, 10- and 12-fold rotational symmetry. Atom positions are identified using clustering of feature descriptors. Subsequent nearest-neighbor analysis and border following on the interatomic connections deliver the tiling. Support vector machines further increase the quality of the results, reaching an accuracy consistent with those reported in the literature. A statistical analysis of the results is performed. The code is now part of the open-source package AiSurf.
\end{abstract}

\keywords{quasicrystals, pattern recognition, machine learning, STM, STEM}

\maketitle

%\tableofcontents

%%%MAIN TEXT%%%%
\section{Introduction}
Quasicrystals are a class of aperiodic materials showing long-range ordering. While quasicrystalline patterns have appeared in various early artworks and architectural designs~\cite{islamart}, the first experimental observation of a ten-fold electron diffraction pattern, revealing non-periodic tilings of the plane, was reported by Dan Shechtman {\em et al.} in 1984~\cite{shechtman1984QC}. 
%Since the first publication about quasicrystals in 1984~\cite{shechtman1984QC}, a vast interest of the scientific community on the topic has been reported, testifying to a revolution in solid-state physics. 
Since then, quasicrystalline systems have been extensively studied for their natural beauty and unusual characteristics, leading to the discovery of several remarkable physical properties~\cite{louzguine2008formation}. Some quasicrystals have been reported to show high electrical resistivity~\cite{kashimoto2000resitivity}, and can be suitable for hydrogen storage~\cite{takasaki2006hydrogen, gibbons2004hydrogen}; the lack of translational symmetry makes the quasicrystals usually hard and brittle at room temperature~\cite{edagawa2001dislocations}. Recent studies report bulk superconductivity~\cite{takemori2024superconducting, tokumoto2024superconductivity}, nematic superconductivity~\cite{liu2024nematicSC} and asymmetric phonon intensities~\cite{matsuura2024phononQC}.

It is well known that crystalline lattices are only compatible with 1-, 4- and 6-fold rotational symmetries~\cite{hoffmann2020crystallography}, and their ideal structure can be described in terms of a single unit cell. Quasicrystalline structures, however, do not possess the periodicity paradigm of classical crystallography as 5-, 8-, 10- and 12-fold rotational symmetries are allowed~\cite{stadnik2012QC}: they show Bragg spots in the diffraction pattern, but no translational periodicity~\cite{yamamoto1996crystallography}. This poses major challenges for the crystallographical description of such structures, but also attracts the attention of a vast scientific community that tries to identify structure-property correlations~\cite{dolinvsek2012electrical}. Not describable using single unit cells, quasiperiodic patterns can be generated by packings, coverings, and tilings~\cite{steurer2008fascinating}. Recent studies show promising steps towards a more generalized, formal description of the tilings~\cite{imperor2024STR, imperor2021ST}.
These tilings, such as the renowned Penrose tiling, consist of primitive tiles that can be arranged into an aperiodic pattern by following matching rules~\cite{grunbaum1987tilings}. They can be interpreted as the projections of lattices from a higher-dimensional space~\cite{yamamoto1996crystallography}.
In recent years, the first oxide quasicrystal (OQC) has been observed~\cite{forster2013quasicrystalline, forster2020quasicrystals} and other studies followed on that direction~\cite{forster2016observation, zollner2020BTOqc, wuhrl2023quasicrystal, merchan2022oqcSTO}, hinting for a promising, unexplored area of surface science.
\\
Atomically resolved images can reveal several mechanisms and features encompassing crystallographic information, domain orders, molecular adsorption, and distribution of defects.
During the last decade, the interest in automated analysis and pattern recognition for microscopy images has significantly grown~\cite{kalinin2023automated, kalinin2022MLprimer}.
Several different approaches are proposed in the literature to extract information from these images. WSXM~\cite{horcas2007wsxm} and ImageJ~\cite{collins2007imagej} are popular choices for manual analysis. Machine learning has seen an exponential increase in popularity in recent years, and its applications in the field of microscopy are now several~\cite{kalinin2023probe, kalinin2023ML, pregowskascanning}, such as convolutional neural networks~\cite{patrick2023grainCNN, madsen2018deep, leitherer2023BNN, ziletti2018insightful} and graph neural networks~\cite{li2023GNN}.
Unsupervised methods also saw an increase in popularity, starting from clustering algorithms~\cite{corrias2023automated}, variational autoencoders~\cite{ziatdinov2023simplicity, biswas2023vae}, support vector machines~\cite{guo2021defect}, and others~\cite{somnath2018feature}.
Machine learning and data-driven approaches have been applied to identify, predict, and design quasicrystalline structures~\cite{uryu2024mlQC, liu2021mlQC, Chang2023, Fujita2024}. However, to the best of our knowledge, no publicly available methods have been specifically designed to extract quasicrystalline patterns from microscopy images.
\\
In this work, we introduce a pattern
recognition tool tailored for quasicrystalline surfaces, based on supervised and unsupervised machine learning.
Its performance is tested on atomically resolved scanning tunneling microscopy (STM) and High-angle annular dark-field imaging (HAADF) images of polygonal quasicrystal and approximant surfaces.
The workflow is presented in the next section, followed by results and discussions. Data, code, and documentation are available as part of the open-source AiSurf Python package~\cite{corrias2023automated}.
\label{sec:intro}

\section{Methods}
\begin{figure*}
    \centering
    \includegraphics[width=2\columnwidth]{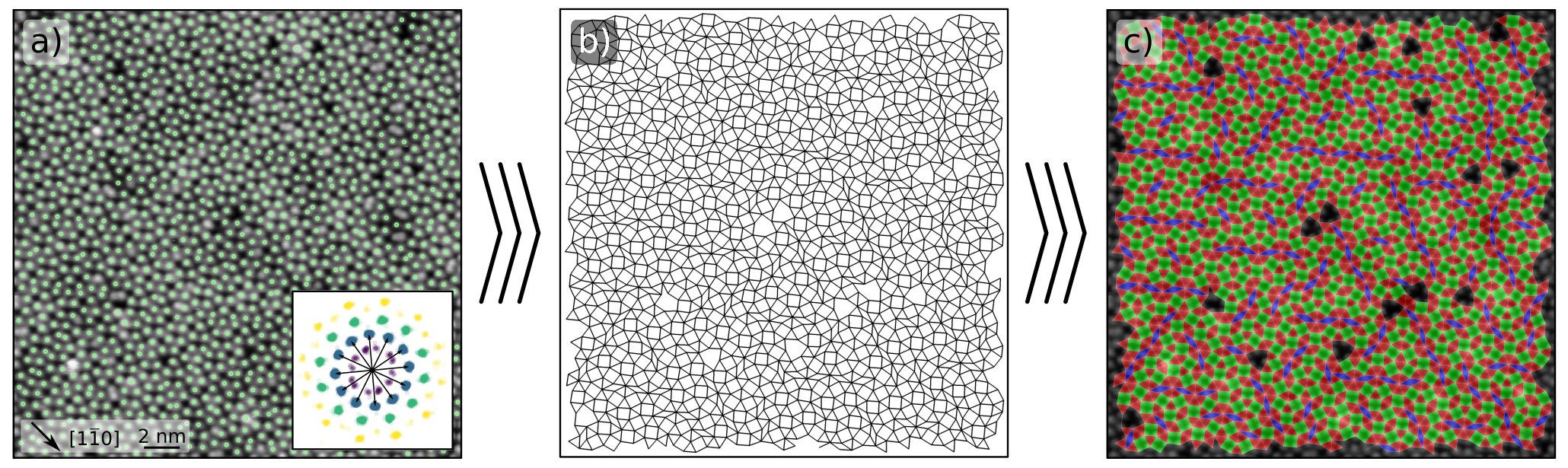}
% \begin{subfigure}{.3\textwidth}
%     \centering
%     \includegraphics[width=\textwidth]{figures/atom_positions_paper.png}
%     \caption{}
%     \label{fig1_a}
% \end{subfigure}
% \begin{subfigure}{.3\textwidth}
%     \centering
%     \includegraphics[width=\textwidth]{figures/1_contours.png}
%     \caption{}
%     \label{fig1_b}
% \end{subfigure}
% \begin{subfigure}{.3\textwidth}
%     \centering
%     \includegraphics[width=\textwidth]{figures/1_filled_tiling_run1_inlay.png}
%     \caption{}
%     \label{fig1_c}
% \end{subfigure}
\caption{Key steps in the quasiperiodic pattern recognition process on an STM image of a 12-fold rotationally symmetric BaTiO$_3$ quasicrystal on Pt(111). Adapted with permission.~\cite{forster2013quasicrystalline} Copyright 2025, Nature. (a) Detected atomic keypoints shown by green circles with radii indicating the reduced keypoint size. Bottom right: nearest-neighbor distribution with $k=7$, color-coded by cluster labels with different distance from the origin. The black arrows represent the symmetry vectors pointing to the 12 clusters of relative nearest-neighbor locations in the dominant distance cluster (blue). (b) Tiling outline created by connecting each atomic keypoint center to the center of all its neighbors in the dominant distance cluster. A border-following algorithm is applied to this image to extract contours. (c) Final tiling after completing missing point search in "open" contours. Tile colors are assigned based on their internal angles. Holes in the tiling stem from vacancies in the quasicrystal.
}
\label{fig1}
\end{figure*}

Pattern recognition utilizes image feature extraction as a means of finding atomic positions. To this end, the Scale Invariant Feature Transform (SIFT)~\cite{lowe2004distinctive} algorithm is employed. By repeatedly convolving atomically resolved images with a Gaussian function with increasing $\sigma$ and taking the difference between two consecutive convolutions, a bandpass filter is applied, acting as a feature enhancer. From these stacks of Difference of Gaussian (DoG) images, split by limiting the increase of $\sigma$ to double its initial value and downsampling before continuing the convolutions, extremum points are extracted as the features of the images, referred to as "keypoints". Although the SIFT parameters require some tuning for optimal results depending on the image quality and resolution, the default values are set to ensure meaningful initial results.
\\
Filtering out keypoints outside of a set range around the median keypoint size discards the first batch of non-atomic keypoints. The remainder, atomic keypoints and non-atomic keypoints of atomic size, are resized to a fraction of their diameter such that the descriptor, computed in succession, captures the central gradient. The descriptor is a histogram of the local gradient orientations and magnitudes characterizing a keypoint, often referred to as "histogram of oriented gradients" (HOG).
\\
The keypoints are split through agglomerative clustering of the descriptors with the cluster amount set to 2, so that positive and negative central gradients are seen as a distinct class of descriptors each. Subsequent $k$-nearest-neighbor (kNN) analysis determines which of the two keypoint clusters contains the atomic keypoints. For each keypoint cluster, after calculating the $k$ nearest-neighbors, with $k$ set between $5$ and $10$, the absolute distances from the origin point to the neighbors are clustered, with the silhouette score~\cite{Rousseeuw1987silhouette} used to determine the optimal cluster amount. The result is what we refer to as a nearest-neighbor (NN) plot. The atomic keypoints will show the highest distance clustering silhouette score due to relatively discrete nearest-neighbor distance steps. Due to the aforementioned choice of $k$ in the kNN calculation, keypoints in the dominant distance cluster, that is, the distance cluster with the most members, will contain the next vertex points in the tiling. Therefore, by clustering the relative keypoint positions of members in this dominant distance cluster and using silhouette scoring with the cluster amounts limited to 8, 10, or 12, the rotational symmetry is determined. The median location of these vertex point clusters indicates the symmetry vectors of the underlying tiling. An example result of this process can be seen in figure \ref{fig1}\textcolor{red}{a}, which shows an STM image of a dodecagonal BaTiO$_3$ quasicrystal grown on Pt(111) with green circles indicating positions and size of the atomic keypoints after reduction. The bottom right inset shows the nearest-neighbor distribution for the atomic keypoints with $k=7$, colored by a distance cluster label with symmetry vectors pointing to the dominant distance cluster.
\\
A binary image is created where the lines connect an atomic keypoint to each of its neighbors in the dominant distance cluster, as shown in Fig.\ref{fig1}\textcolor{red}{b}. Using the border-following algorithm proposed by Suzuki et al.~\cite{suzuki1985topological}, an array of contours containing the individual tile outlines is created. Missing feature detections through SIFT, especially for two or more atoms in close proximity, leads to some contours in the array being "open" i.e. containing potential missed atom positions.
Open contours are used to search for missing points after completion of the initial tiling recognition. The contour array is split into tiles and open contours by applying a contour shape constraint followed by clustering. The constraint is based on observed quasicrystal tilings, which are generally composed of triangles, squares, rhombs~\cite{imperor2024STR} or hexagons~\cite{wang2020decagonal}. Therefore, any contour whose corner numbers do not conform with those shapes is immediately classified as "open". 
For each of the remaining contours, a feature vector made up of their inner angles is created. Clustering these feature vectors splits the tiles into their shape groups and allows to move any remaining outliers (contours in clusters with relatively few members) to the array of open contours. 
\\
In preparation for the missing point search, a classifier is trained on the clustering results of the keypoint descriptors. For this purpose a support vector machine (SVM)~\cite{cortes1995support} was chosen, a supervised learning method that searches for the hyperplane with maximal margin that separates classes in labeled training data. Subsequently, symmetry vectors obtained through the nearest-neighbor analysis are added to each corner keypoint in the "open" contours. The resulting locations inside the "open" contours are clustered together, and a keypoint with a descriptor is created at each cluster's median location. The SVM decides whether the newly formed candidate keypoint is on top of an atom in the original image. The nearest-neighbor analysis and tile identification are performed again after a complete search to update the tiling. Images with large open areas may need multiple runs of the missing point search process to be filled. The final output of the pattern recognition process for BaTiO$_3$ with tiles colored by their feature vector cluster labels is shown in figure \ref{fig1}\textcolor{red}{c}. A statistical analysis of the tile frequency, orientational and angular distribution can be performed, as discussed later.

\section{Results \& discussion}
\begin{figure*}[t]
    \centering
    \includegraphics[width=2\columnwidth]{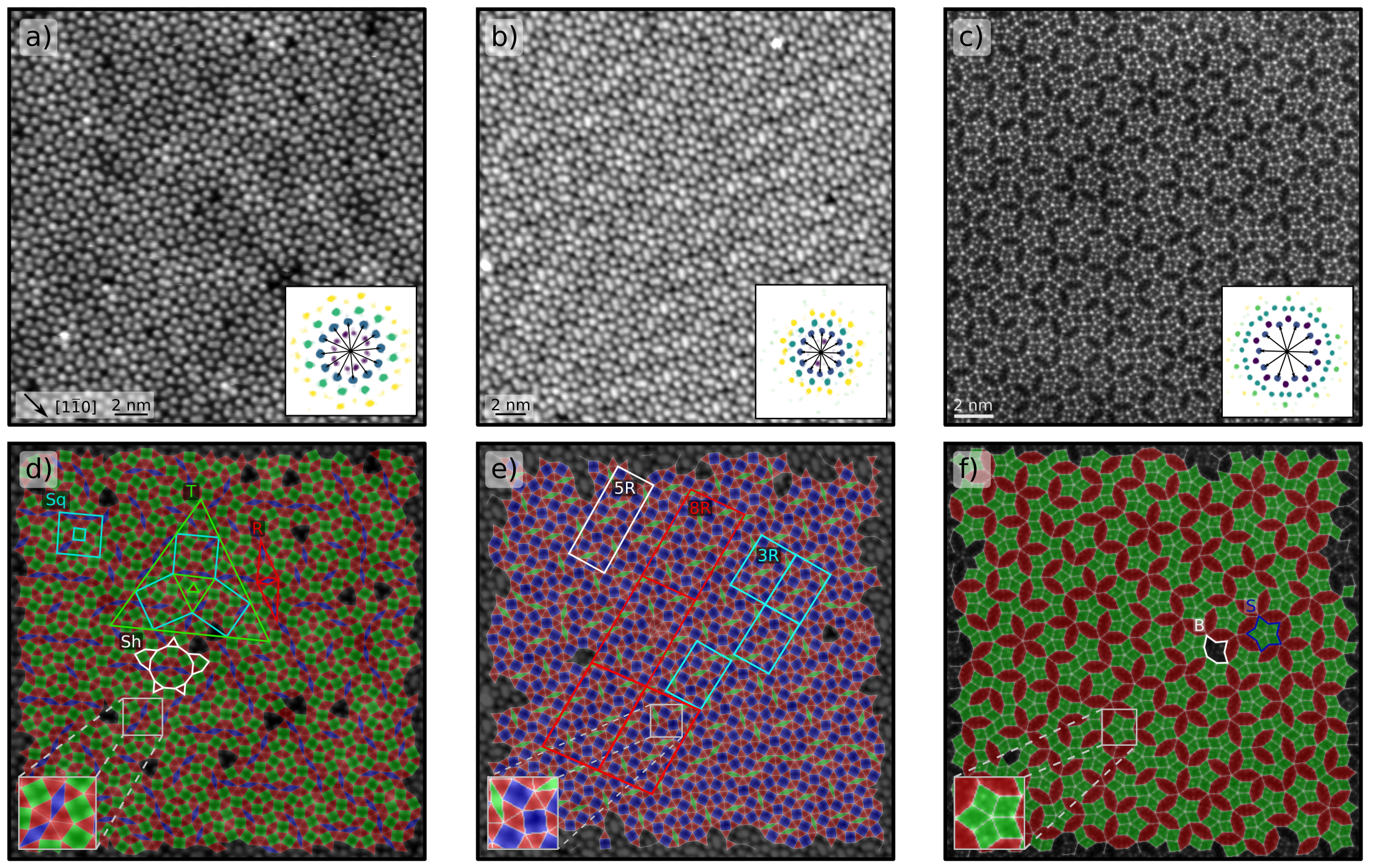}
    \caption{
    a) STM image of BaTiO$_3$-derived OQC on Pt(111).
    Adapted with permission.~\cite{forster2013quasicrystalline} Copyright 2025, Nature.
    Nearest-neighbor (NN) plot of the atomic coordinates is inserted as an inset, highlighting the 12-fold symmetry of the system.
    b) STM image of Ba-Ti-O layer on Pd(111). Adapted with permission~\cite{wuhrl2023quasicrystal}, Copyright 2025, APS. The NN plot reveals a 12-fold rotational symmetry.
    c) HAADF-STEM image of a quasicrystal in a Zn$_{58}$Mg$_{40}$Y$_2$ alloy. Adapted under terms of the CC-BY license.~\cite{wang2020decagonal}, Copyright 2025, IUCr. Image taken along the tenfold axis. The 10-fold rotational symmetry is highlighted by the nearest-neighbor plot.
    d) Highlighted patterns from the tiling performed on the image in panel \textit{a}: among them we have the deflation/inflation of a triangle (T), a square (Sq), and a rhomb (R). A so-called "ship tile" (Sh) is also highlighted. The bottom-left inset shows a magnification of a small region, the bottom-right one shows the extracted tiles, alongside their relative frequency and internal angles.
    e) Tiling obtained from the image in panel \textit{b}. Green rhombs, red triangles, and blue squares are collected in locally quasiperiodic patterns with 12-fold rotational symmetry. Some \textit{3R}, \textit{5R} and \textit{8R} approximant cells are highlighted. The bottom-left inset shows a magnification of a small region.
    f) Tiling obtained from the image in panel \textit{c}, consisting of elongated hexagons (red) and rhombs (green). A boat tile (B) and a star tile (S) are highlighted. The bottom-left inset shows a magnified star tile.
    }
    \label{fig:other_tiles}
\end{figure*}

In this section, we present the results of the tiling obtained from   the tested systems. 
Figs.~\ref{fig:other_tiles}\textcolor{red}{a-c} show the atomically resolved images used for this study: (a) BaTiO$_3$ on a Pt(111), (b) Ba-Ti-O grown on Pd(111), and (c) Zn$_{58}$Mg$_{40}$Y$_2$ alloy.
Their nearest-neighbor (NN) plot, displaying the relative atomic positions, is inserted as a subpanel, from which one observes a regularity in the atomic coordinates. After calculating the symmetry vectors as explained in the Methods section, we evince that Figs.~\ref{fig:other_tiles}\textcolor{red}{a, b} show a 12-fold rotational symmetry, while Fig.~\ref{fig:other_tiles}\textcolor{red}{c} a 10-fold one.
Fig.~\ref{fig:other_tiles}\textcolor{red}{a} shows a 2D OQC obtained from BaTiO$_3$ on a Pt(111) surface, the same system shown in Fig.~\ref{fig1}.
Fig.~\ref{fig:other_tiles}\textcolor{red}{b} shows an STM image of an OQC on Ba-Ti-O grown on Pd(111), also discussed in W{\"u}hrl et al.~\cite{wuhrl2023quasicrystal}.
Fig.~\ref{fig:other_tiles}\textcolor{red}{c} shows a HAADF-STEM image of a quasicrystal in a Zn$_{58}$Mg$_{40}$Y$_2$ alloy, as discussed in Wang et al.~\cite{wang2020decagonal}.

Fig.~\ref{fig:other_tiles}\textcolor{red}{d} shows the tiling results obtained from Fig.~\ref{fig:other_tiles}\textcolor{red}{a}. The tiling is composed of three elements: a triangle, a square, and a rhomb. Excluding boundaries and vacancies, we observe that the entire surface has been accurately covered by tiles. From the nearest-neighbor plot shown in the inset, we observe that the system possesses dodecagonal symmetry. 
As discussed in~\cite{schenk2019full}, the atomic arrangement can be modeled with a Niizeki-Gähler tiling (NGT)~\cite{gahler1988quasicrystalline, niizeki1987dodecagonal}.
Although the tiling does not fully conform to an NGT along the whole surface, some prominent structures can be highlighted. Such structures are present multiple times in the image, but only one of each type has been highlighted for demonstrative purposes. Our data shown in  Figure~\ref{fig:other_tiles}\textcolor{red}{d} exhibit
the so-called "ship/turtle tile" pattern~\cite{liao2013single} ('Sh' in Fig.~\ref{fig:other_tiles}\textcolor{red}{d}, demarkated by white lines) 
already reported in other works. The "ship/turtle tile" pattern
% so-called "ship/turtle tile" pattern~\cite{liao2013single} ('Sh' in Fig.~\ref{fig:other_tiles}\textcolor{red}{d}) is highlighted with white lines: 
consists of 19 triangles, 7 squares, and 2 rhombs, differing slightly from the one reported in the literature (which contains 20 triangles, 7 squares, and 2 rhombs). This difference arises because the "tails" of the ship are composed of three adjacent triangles instead of two.
The central part of this tile is a dodecagon formed by 2 rhombs, 5 squares, and 12 triangles; this combination is repeated several times in the image, as highlighted in~\cite{forster2013quasicrystalline}.
Due to the recursive nature of NGT~\cite{liao2013single}, unit tiles can be deflated and inflated, meaning that larger tiles can be obtained from a composition of smaller ones, as highlighted in Fig.~\ref{fig:other_tiles}\textcolor{red}{d}: from an ideal NGT, a larger triangle can be decomposed into 7 smaller triangles and 3 squares, a larger square into 16 triangles, 5 squares and 4 rhombs, and a larger rhomb into 8 triangles, 2 squares and 3 rhombs.
Our analysis finds a perfect match with the two recursions of the triangle ('T') and the recursion of the rhomb ('R'), whereas the highlighted square ('Sq') and the ones composing the larger triangle display a slightly different composition with respect to each other and the ideal NGT.

Fig.~\ref{fig:other_tiles}\textcolor{red}{e} shows the tiling results of Fig.~\ref{fig:other_tiles}\textcolor{red}{b}. This surface shows a variety of quasicrystal approximants, as discussed in~\cite{wuhrl2023quasicrystal}. Approximant tilings can be interpreted as the projection of a lattice from a higher dimension at an angle that approximates an irrational number~\cite{macia2019chemical}. They thus have a rotational symmetry similar to that of quasicrystals but show periodic traits. A locally quasiperiodic pattern will show multiple repetitions across the surface. Due to their rotational symmetry and local quasiperiodicity, they can be analyzed with our pattern recognition tool. 
Like in Fig.~\ref{fig:other_tiles}\textcolor{red}{d}, this surface can be modeled with an NGT, consisting of triangles, squares, and rhombs with a fixed relative ratio, compatible with the ideal case. Except for a small number of missing keypoints and a misclassification error near the center of the image (where a more complex pattern was decomposed into small triangles), the program was able to perform the tiling correctly. We highlight some of the three different approximant cells discussed in the original paper: \textit{3R}, \textit{5R}, \textit{8R}, named after the number of rhombs ('R') present in each cell. In particular, \textit{8R} is the largest quasicrystal approximant reported to date~\cite{wuhrl2023quasicrystal}.
An ideal \textit{3R} contains 24 triangles, 9 squares, and 3 rhombs, as in our case.
An ideal \textit{5R} is composed of 40 triangles, 15 squares, and 5 rhombuses, as in our highlighted case.
An ideal \textit{8R} contains 60 triangles, 22 squares, and 8 rhombs. This is also compatible with the cells we highlighted, except for the region where the misclassification error occurred. This supports the consistency of our tiling in this system. 

Fig.~\ref{fig:other_tiles}\textcolor{red}{f} shows the result of the tiling of Fig.~\ref{fig:other_tiles}\textcolor{red}{c}. This surface can be modeled with rhombic and hexagonal tiles, and from the nearest-neighbor plot, we observe that the system has decagonal symmetry. The surface is mostly constituted by rhombs, which are frequently displayed together in groups of five to form "star" motifs, labeled with 'S' in Fig.~\ref{fig:other_tiles}\textcolor{red}{f}. In the left part of the image, the atomic arrangement is less regular, leading to open shapes made of rhombs referred to as "zigzag" in the literature~\cite{wang2020decagonal}. A single boat tile 'B', as referenced in literature, is also present in the image. \\
The results we have shown are highly consistent with the original works, thus proving the reliability of this code and suggesting its applications to other systems.

\begin{figure}
    \centering
    \includegraphics[width=0.9\columnwidth]{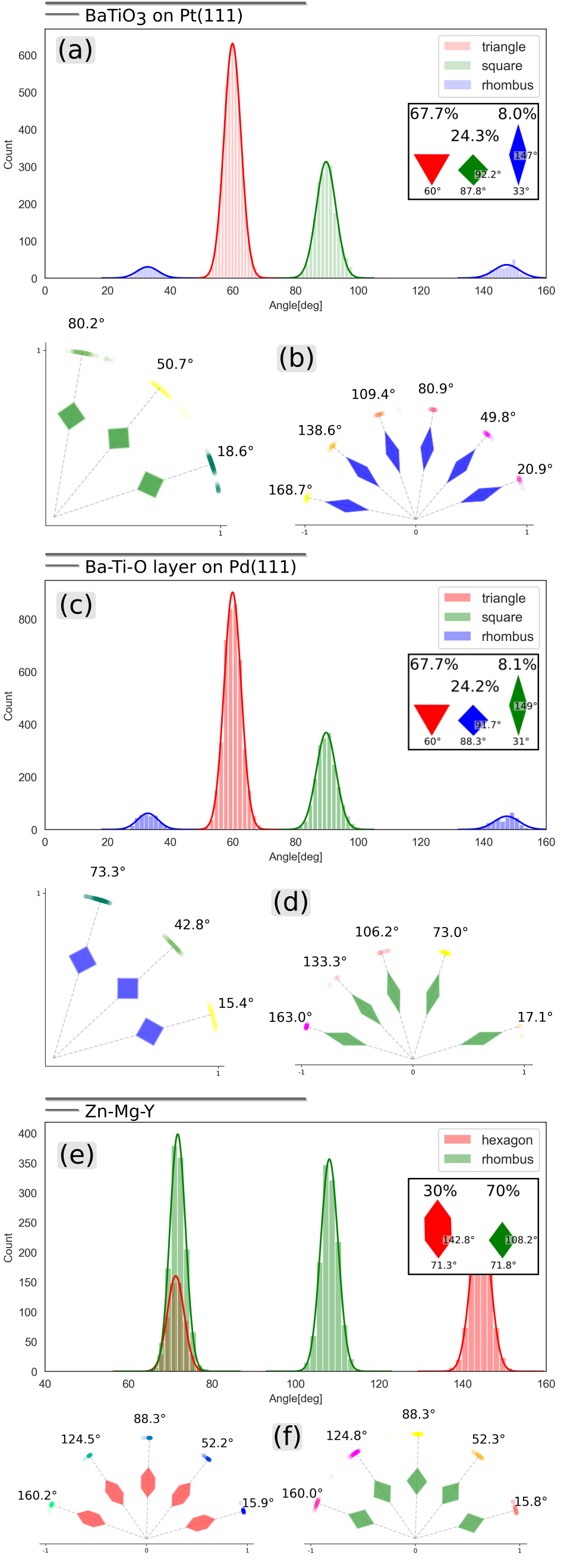}
\caption{Statistical analysis of the tiling shown in Fig.~\ref{fig:other_tiles}.
(a, c, e) distribution of the internal angles of each tile, with insets displaying the average appearance of each tile, its internal angles, and frequency.
(b, d) orientational distribution of squares (left) and rhombs (right) from Figs.~\ref{fig:other_tiles}\textcolor{red}{d, e}, respectively, with median values highlighted. (f) Orientational distribution of hexagons (left) and rhombs (right) from Fig.~\ref{fig:other_tiles}\textcolor{red}{f}.
}
\label{fig:statistics}
\end{figure}

Lastly, we present the statistical analysis of the tilings generated by the program, following the methodology established by Schenk \emph{et al.}~\cite{schenk2019full}. For each image presented, plots related to the internal angles, orientations, and tile frequency have been obtained. The tilings, these plots and the numerical values extracted from them allow the users to perform both a qualitative and quantitative analysis of the systems under study.
Fig.~\ref{fig:statistics}\textcolor{red}{a} shows the distribution of the internal angles related to the tiling in Fig.~\ref{fig:other_tiles}\textcolor{red}{d}. As expected, the angles are spread around a central value, approximately following Gaussian distributions. The curves related to triangles and squares peak at 60° and 90°, the ones for rhombs at 147° and 33°. The Full Width at Half Maximum (FWHM) of each curve is comparable with the others: 6.5° for triangles, 7.9° for squares, 7.1° and 8.8° for rhombs (from left to right). The inset on the right of the histogram shows the frequency of the tiles appearing in the image, along with their average internal angles.
The relative tile ratio conforms with the ideal NGT, where the fraction of triangles, squares, and rhombs with respect to the squares is (2.72, 1, 0.37)~\cite{schenk2019full}. In our case we have (2.79, 1, 0.33), close to the predicted values.
Except for the triangles, the average internal angles of each tile show slight deviations with respect to the ideal values. The squares have an average internal angle of 87.8° and 92.2°, while the rhombs 33° and 147°. These values are associated with local distortions of the atomic arrangement. \\
Fig.~\ref{fig:statistics}\textcolor{red}{b} shows the rotational distribution of squares (left) and rhombs (right). The squares are oriented around the median values shown in the figure, although two peaks are present for each orientation; this hints at a systematic distortion of the tiles, not noticeable by the naked eye, that can be further investigated. Apart from some outliers, the orientational distribution of the rhombs does not present particular features to comment on.
\\
Figs.~\ref{fig:statistics}\textcolor{red}{c, d} show the analysis of Fig.~\ref{fig:other_tiles}\textcolor{red}{e}. The histogram in Fig.~\ref{fig:statistics}\textcolor{red}{c}, as in the previous case, shows distributions of internal angles approximable to Gaussians.
From left to right, with the colors indicating the respective tiles, we have peaks at 30°, 60°, 90° and 150°, and FWHM at 4.5°, 7.0°, 
5.7°, 5.0°.
The inset on the left displays slight average distortions of squares and rhombs, less pronounced than in Fig.~\ref{fig:statistics}\textcolor{red}{a}. The tile frequency is also well compatible with an ideal NGT: in this case we have (2.8, 1, 0.33). Fig.~\ref{fig:statistics}\textcolor{red}{d} shows the orientational distributions of squares (left) and rhombs (right), with median values highlighted. It is worth noticing that one of the possible orientations of the rhombs is missing, and we have not identified an explanation for this oversight.
\\
The statistical analysis of Fig.~\ref{fig:other_tiles}\textcolor{red}{f} is presented in Figs.~\ref{fig:statistics}\textcolor{red}{e, f}. The tiling of this system is described by hexagons and rhombs. The internal angles of each tile follow a Gaussian distribution, with two angles of hexagons and rhombs overlapping. The Gaussian fits for hexagons have peaks at 71.3° and 144.3°, for rhombs at 71.7° and 108.0°, with respective FWHM at 4.5°, 5.4°, 8.2° and 7.2°.
There is a 30:70 ratio between hexagons and rhombs. Like in the previous cases, the local distortions of the tilings are shown by the orientational distribution in Fig.~\ref{fig:statistics}\textcolor{red}{f} of hexagons (left) and rhombs (right).
% \label{sec:results_disc}

\section{Conclusions}
% \label{sec:conclusions}
An automated pattern recognition tool for atomically resolved surface images of polygonal quasicrystals has been outlined, implemented, and tested. Atom positions are identified using the feature recognition algorithm SIFT coupled with agglomerative clustering and nearest-neighbor analysis. A subsequent border-following algorithm on connected atom positions is applied to find the individual tile components. The shortcomings of the SIFT feature recognition algorithm are corrected by using the symmetry vectors that characterize the tiling and an SVM trained on the classified initial features to search for missing atoms. \newline
The workflow has been tested on an STM image of a dodecagonal BaTiO$_3$ quasicrystal, an HAADF image of a decagonal quasicrystal in a Zn$_{58}$Mg$_{40}$Y$_2$ alloy and an STM image of a Ba-Ti-O quasicrystal approximants with 12-fold rotational symmetry; all tests had a positive outcome, showing the high accuracy of the workflow.
All tests are highly consistent with the tiling performed in previous studies, proving the reliability of the method. The algorithm is conceptualized to take 8-fold rotational symmetry into account as well, as it serves as a recognition tool for all types of polygonal quasicrystals.
A statistical analysis of the results has been performed, which is useful for spotting and quantifying local distortions.
\newline
This tool provides a user-friendly yet precise method to extract and visualize tiling patterns from atomically resolved microscopy images. Furthermore, its general applicability makes it a valuable resource for analyzing other quasicrystalline systems beyond those presented in this manuscript. A public version of the tool is provided as an extension of the AiSurf Python package.

\section*{Author contributions}
TKK and MC contributed equally to this work.
TKK developed and implemented the workflow with assistance from MC, and both wrote the article. CF designed and supervised the research and provided revisions to the manuscript.

\section*{Conflicts of interest}
There are no conflicts of interest to declare.

\section*{Data availability}
The AiSurf package, including the code used for this study, can be accessed from \url{https://github.com/QuantumMaterialsModelling/AiSurf-Automated-Identification-of-Surface-images}.

\section*{Acknowledgements}
We kindly thanks Stefan F{\"o}rster for the precious technical discussions on quasicrystals, and for providing the images introduced in Fig.~\ref{fig:other_tiles}\textcolor{red}{a, b}. We thank WZ Wang for sharing the image introduced in Fig.~\ref{fig:other_tiles}\textcolor{red}{c}. \\
This research was funded in whole or in part by the Austrian Science Fund (FWF) 10.55776/F81. For Open Access purposes, the author has applied a CC BY public copyright license to any author-accepted manuscript version arising from this submission.

%%%END OF MAIN TEXT%%%

%%%REFERENCES%%%
\bibliography{biblio}

\end{document}